\documentclass[letter]{aa}
\usepackage{graphicx}
\usepackage{txfonts}
\usepackage{gensymb} 
\usepackage{xspace}
\usepackage{natbib,twoopt}
\usepackage[breaklinks=true]{hyperref} 
\bibpunct{(}{)}{;}{a}{}{,} 
\makeatletter
\newcommandtwoopt{\citeads}[3][][]{\href{http://adsabs.harvard.edu/abs/#3}%
{\def\hyper@linkstart##1##2{}%
\let\hyper@linkend\@empty\citealp[#1][#2]{#3}}}
\newcommandtwoopt{\citepads}[3][][]{\href{http://adsabs.harvard.edu/abs/#3}%
{\def\hyper@linkstart##1##2{}%
\let\hyper@linkend\@empty\citep[#1][#2]{#3}}}
\newcommandtwoopt{\citetads}[3][][]{\href{http://adsabs.harvard.edu/abs/#3}%
{\def\hyper@linkstart##1##2{}%
\let\hyper@linkend\@empty\citet[#1][#2]{#3}}}
\newcommandtwoopt{\citeyearads}[3][][]%
{\href{http://adsabs.harvard.edu/abs/#3}
{\def\hyper@linkstart##1##2{}%
\let\hyper@linkend\@empty\citeyear[#1][#2]{#3}}}
\makeatother
\def\ms{\hbox{m\,s$^{-1}$}}         
\def\cms{\hbox{\,cm\,s$^{-1}$}}       
\def\m2s2{\hbox{\,m$^{2}$\,s$^{-2}$}} 
\def\kms{\hbox{\,km\,s$^{-1}$}}       
\def\vsini{\hbox{$v$\,sin\,$i_{\star}$}}      
\def\Msun{\hbox{$M_{\odot}$}}             

\def\Mjup{\hbox{$\mathrm{M}_{\rm J}$}}

\def\Mearth{\hbox{$\mathrm{M}_{\rm E}$}}
\def\Rearth{\hbox{$\mathrm{R}_{\rm E}$}}

\def\ten[#1]{$\;\times 10^{#1}$}

\def\logg{$\log g$}

\newcommand{\e}[1]{{\times10^{#1}}}

\newcommand{\Rnom}{\hbox{$\mathcal{R}^{\rm N}_{\odot}$}} 

\newcommand{\GMnom}{\hbox{$\mathcal{(GM)}^{\rm N}_{\odot}$}}

\newcommand{\Renom}{\hbox{$\mathcal{R}^{\rm N}_{e \rm E}$}}

\newcommand{\GMenom}{\hbox{$\mathcal{(GM)}^{\rm N}_{\rm E}$}}

\newcommand{\RJnom}{\hbox{$\mathcal{R}^{\rm N}_{e \rm J}$}}

\newcommand{\GMJnom}{\hbox{$\mathcal{(GM)}^{\rm N}_{\rm J}$}}

\newcommand{\reb}{{\sc \tt REBOUND}\xspace}
\newcommand{\whf}{{\sc \tt WHFast}\xspace}
\newcommand{\emcee}{{\sc \tt emcee}\xspace}

\begin{document} 

     \title{Absolute densities, masses, and radii of the WASP-47 system determined dynamically}

   \author{J.M.~Almenara\inst{\ref{geneva},\ref{grenoble},\ref{lam}}
      \and R.F.~D\'{i}az\inst{\ref{geneva}, \ref{iafe}}
      \and X.~Bonfils\inst{\ref{grenoble}}
      \and S.~Udry\inst{\ref{geneva}}
      }

      \institute{
        Observatoire de Gen\`eve, Département d’Astronomie, Universit\'e de Gen\`eve, Chemin des Maillettes 51, 1290 Versoix, Switzerland\label{geneva}
        \and Univ. Grenoble Alpes, CNRS, IPAG, F-38000 Grenoble, France\label{grenoble}
        \and Aix Marseille Universit\'e, CNRS, LAM (Laboratoire d'Astrophysique de Marseille) UMR 7326, 13388, Marseille, France\label{lam}
        \and Instituto de Astronom\'ia y F\'isica del Espacio (CONICET-UBA), C.C. 67, Sucursal 28, C1428EHA, Buenos Aires, Argentina\label{iafe}
        }
      \date{}

      \date{}

 
  \abstract
      {
        We present a self-consistent modelling of the available light curve and radial velocity data of WASP-47 that takes into account the gravitational interactions between all known bodies in the system. The joint analysis of light curve and radial velocity data in a multi-planetary system  allows deriving absolute densities, radii, and masses without the use of theoretical stellar models. For WASP-47 the precision is limited by the reduced dynamical information that is due to the short time span of the K2 light curve. We achieve a precision of around 22\% for the radii of the star and the transiting planets, between 40\% and 60\% for their masses, and between 1.5\% and 38\% for their densities. All values agree with previously reported measurements. When theoretical stellar models are included, the system parameters are determined with a precision that exceeds that achieved by previous studies, thanks to the self-consistent modelling of light curve and radial velocity data.        
      }
   \authorrunning{J.M. Almenara et al.}
   \titlerunning{WASP-47}

   \maketitle
%

\section{Introduction}

The WASP-47 planetary system is composed of at least four planets. The first planet to be discovered was the transiting hot Jupiter WASP-47~b in a 4.16~days orbit \citep{hellier2012}, whose mass was determined by radial velocity measurements. Long-term radial velocity follow-up detected a second massive long-period ($\sim$570~days) planet, WASP-47~c \citep{neveu2016}, which is not known to transit the star. Observations from the K2 mission \citep{howell2014} that lasted 69~days allowed detecting two additional transiting planets: a Neptune-sized planet on a 9-day orbit (planet d), and a 1.8~\Rearth\ planet on a 0.8-day orbit (planet e) \citep{becker2015}. WASP-47 is the only system known to date with a close-in planet inside the orbit of a hot Jupiter and an external massive companion; a benchmark for planetary formation and migration studies.

The transit-timing variations (TTV) detected in the K2 light curve enabled \citet{becker2015} to determine the mass of planet~d, and to set an upper limit to the mass of planet~e. In addition, they produced an independent measurement of the mass of planet~b that agrees with the mass from radial velocity. \citet{dai2015} monitored the system with high-precision radial velocity and detected the Doppler signature of planet~e. They also obtained a marginal detection of the mass of planet~d, and newly determined the mass of planet~b. In all cases, the results of \citet{dai2015}  agree with the TTV analysis of \citet{becker2015}. \citet{sanchis2015} detected the Rossiter-Mclaughlin effect of planet~b, obtaining a low projected stellar obliquity that is compatible with an aligned orbit.

In this letter we perform the first joint analysis of the available photometric light curves and radial velocity measurements, modelling self-consistently the gravitational interactions between the planets. In line with the ideas introduced in \citet{almenara2015}, we continue to explore the ability of this type of modelling to provide absolute measurements of masses, radii, and densities in multi-planetary systems. Additionally, the self-consistent modelling of WASP-47 improves the determination of the system parameters when we add stellar models.

\section{Observations}

\subsection{Light curves}

We employed the K2 light curve reduced by \citet{becker2015}, obtained during Campaign 3, with a time-span of 69~days, a time cadence of 1~minute, and 350~ppm precision per data-point. This is the data set that contains most of the information on the dynamical interaction between the planets. In addition, we included a high-precision transit of WASP-47~b obtained with EulerCam \citep{lendl2012} that was reported in the original discovery paper by \citet{hellier2012}. The EulerCam light curve was observed using the Gunn r filter and has a temporal sampling of $\sim$77~seconds. We transformed the EulerCam BJD$_{\rm UTC}$ times to BJD$_{\rm TDB}$ \citep{eastman2010} to match the time reference of the K2 data. Only the data spanning three transit durations around each transit mid-time were considered for the analysis, after normalisation with a parabola.

No companion was detected with lucky imaging \citep{wollert2015}, reducing the probability of unaccounted for additional contamination in the photometric aperture.

\subsection{Radial velocities}

We used the published data \citep{hellier2012,neveu2016,dai2015} from the CORALIE \citep{udry2000} and PFS \citep{crane2010} spectrographs. The precision is between 7 and 27~\ms\ for the CORALIE data, and between 2.5 and 4.0~\ms\ for the PFS data. They form a set of 77 radial velocity measurements that we treated as originating in four different instruments, with independent zero-point offsets: the CORALIE measurements presented by \citet{hellier2012} (CORALIE 2012), those presented by \citet{neveu2016} and performed before its upgrade in 2014 (CORALIE 2016a), those performed afterwards (CORALIE 2016b), and the PFS measurements from \citet{dai2015}.

\section{Analysis}

\begin{figure}
\hspace{0.25cm}  \includegraphics[height=6cm]{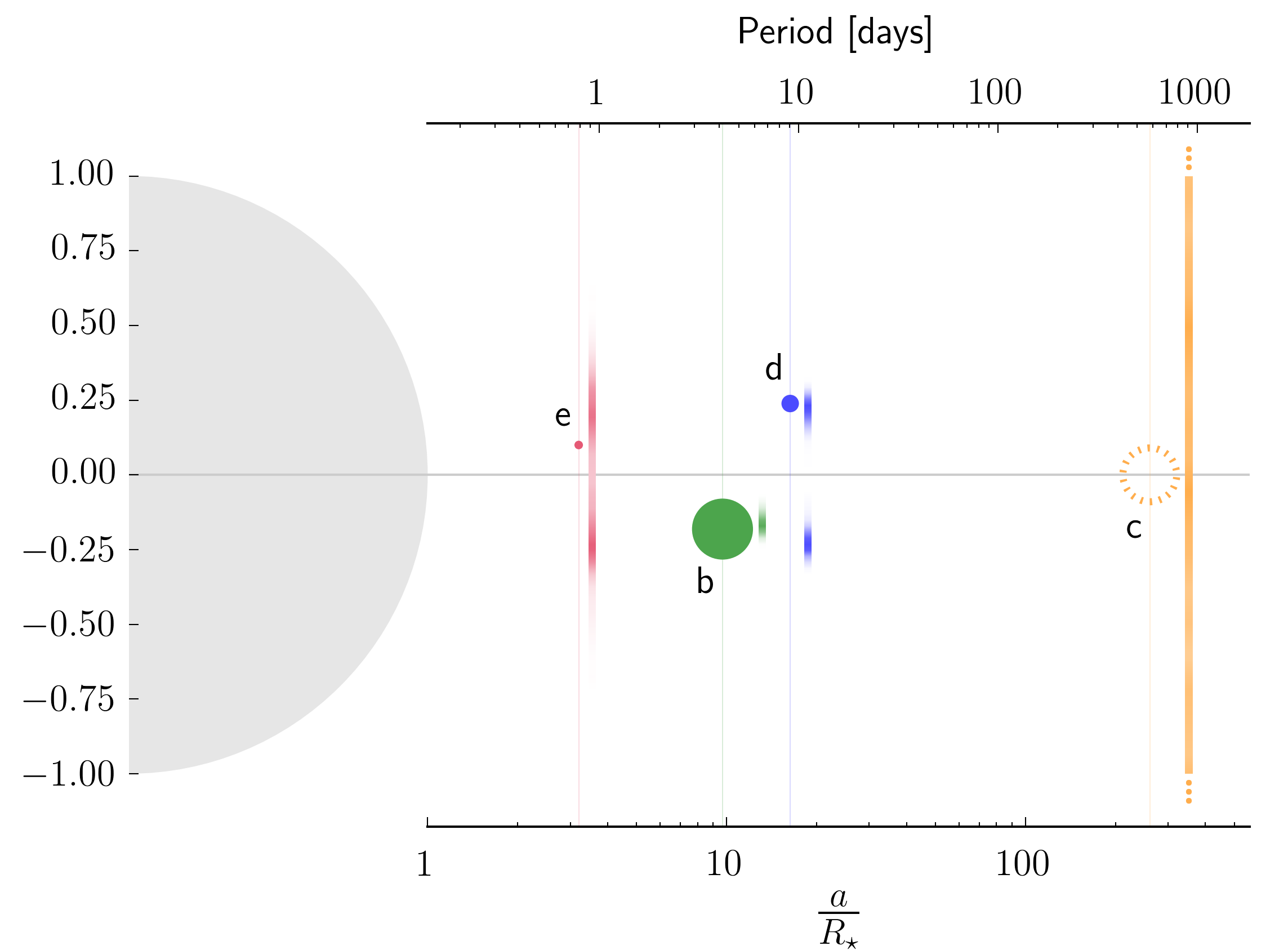}  \caption{Schematic view of the WASP-47 system. The sizes of the star and the planets are to scale, and the distances from the surface of the star are in logarithmic scale. The vertical positions of the filled circles representing the planets correspond to maximum a posteriori estimates of the impact parameters, and the shaded areas (offset in the x direction for clarity) represent the corresponding posterior distributions. The distribution of the planet~c is much larger and is truncated in this figure; its size is unknown.}
  \label{fig.configurations}
\end{figure}

The light curve and radial velocity model accounts for the gravitational interactions between all components of the system. The model is described in detail in \citet{almenara2015}. Briefly, the projected velocity of the star is obtained as a direct output of the the n-body integrator. The light-curve model is obtained using the analytical formulae of \citet{mandelagol2002} with a quadratic limb-darkening law \citep{manduca1977}. The projected centre-to-centre distance between the star and the planets is obtained from the n-body integration. In this analysis, we used \reb \citep{rein2012} with the \whf integrator \citep{rein2015}. The integration time step was set to 0.001~days, which results in a maximum error of 8~\cms\ for the radial velocity model. The combined error in the photometric model is smaller than 6~ppm per observed data point because of the discrete n-body time step and the oversampling factor of 10 \citep{kipping2010}.

The n-body integrator was initialized with the positions and velocities of the system bodies (parametrised by osculating astrocentric asteroidal orbital elements, see Table~\ref{table.results}), at a given reference time, $t_{\mathrm{ref}} = 2\;456\;979.5$~BJD$_{\mathrm{TDB}}$. The reference time $t_{\mathrm{ref}}$ was chosen immediately before the first K2 transit, which is close to the moment containing most of the dynamical information. This means that we used two n-body integrations initialized at $t_{\mathrm{ref}}$, one that was integrated forward in time until the last PFS data point in 2015, and a second one that was integrated backwards in time to the first observation of CORALIE in 2010.
 
The physical parameters of the model are the bodies' masses and radii, the orbital parameters at $t_{\mathrm{ref}}$, and the limb-darkening coefficients for each photometric instrument. To minimise correlations, we parametrised the model using the combinations listed in Table~\ref{table.results}. Additionally, we fitted a normalisation factor for each light curve, an offset value for each radial velocity data set (we assumed zero systemic velocity), and a multiplicative factor to account for additional noise not included in the reported uncertainties (the jitter parameter) for each data set. Although we did not use stellar models, our model implicitly assumes spherical shapes for the star and planets. Under this assumption the model does not depend on the absolute value of the longitude of the ascending node ($\Omega$). Therefore, we fixed $\Omega_{\rm b}$ at $t_{\mathrm{ref}}$ to 180\degree. Finally, there is also a reflection symmetry that allows choosing the hemisphere on which one of the planets transits. We therefore constrained the orbital inclination of planet~b, $i_{\rm b}<90\degree$ (Fig.~\ref{fig.configurations}).

We coupled the dynamical model with the \emcee\ algorithm \citep{goodmanweare2010, emcee} to sample from the posterior distribution of the parameters. The orbital inclination of planet~d exhibits a bimodal distribution with well-separated modes, each corresponding to the planet transiting a different stellar hemisphere. As \emcee\ does not sample well-separated multi-modal distributions efficiently \citep{emcee}, we ran two different \emcee\ runs, one for each mode, and combined the results as explained below. The dynamical interactions are different between these orbital configurations, and it is therefore important to consider them separately, as they could in principle be distinguished. We note that most TTV analyses assume coplanar orbits, as was the case in \citet{becker2015}. Concerning the remaining planets, their inclination distributions are either very wide and do not exhibit a bimodal distribution (planet~c), or the modes are not separate (planet~e). The MCMC algorithm should therefore be able to sample from them correctly. As explained above, the inclination of planet~b was limited to $i_{\rm b}<90\degree$, preventing its bi-modality. 

We used non-informative uniform priors for all 48 \emcee\ jump parameters. We ran \emcee\ with 250~walkers, from a starting point constructed as follows: first, we modelled the radial velocity alone using Keplerian orbits with zero eccentricity for all planets except for planet~c; second, we included the dynamical interactions between the planets, allowed for eccentric orbits, and also modelled the light-curve data. The transit times from \citet{becker2015} were included as an ancillary data set to reduce the algorithm burn-in phase. Then, the final \emcee\ state was used as starting point for the final modelling, for which the transit times were removed. This procedure reduces the burn-in phase in the final photodynamical modelling. For the last step we ran 215\;000 steps of the \emcee\ algorithm and considered the first 190\;000 steps as an additional burn-in period, which was removed for the final inference. The results of the two different orbital configurations described above were combined assuming equal probability for each mode.

\section{Results}
Table~\ref{table.results} lists the maximum a posteriori (MAP) estimate, the median, the 68\% confidence interval (CI), and the 95\% highest density interval (HDI) of the inferred system parameters. Figures~\ref{fig.PH} and \ref{fig.RVt} show the radial velocity measurements and the transit light curves, together with their respective models, Fig.~\ref{fig.TTVs} show the posterior of the TTV. The posterior distributions coming from the two orbital configurations for planet~d are very similar (Fig.~\ref{fig.pyramid}). The analysis seems largely insensitive to the hemisphere on which
planet~d transits, which justifies combining samples from both modes assuming equal probability for each.

The bulk densities are determined with a precision between 1.5\% for the stellar host and 38\% for 1.8-\Rearth\ planet~e. The precision in the stellar density is similar to what can be obtained through asteroseismology \citep{huber2013}. The density of planet~d is 1.8$\pm$0.16~$\mathrm{g\;cm^{-3}}$, which represents a precision of 9\% and constitutes an improvement of almost a factor of 10 with respect to the analysis of \citet{dai2015}. The radii of the star and of all transiting planets are determined with a precision of 22\%. As all planet-to-star radius ratios are very precisely constrained by the transit observations, the knowledge on the absolute radii is dominated by the determination of the stellar radius. Concerning the masses, their posterior distributions are skewed, and it is therefore not trivial to quote a precision value. Using the half-width of the 68.3\%-HDI and its central value, we obtain a precision of between 50\% and 60\% for the star and all planets, except for planet~b, whose absolute mass is known to a precision of around 40\%. We note that this precision is a factor of three lower than the one obtained for the radii. This is surely linked to the fact that as a result of the nature of the gravitational force, the best constrained parameters are the bulk densities of the objects, which involves the third power of the radii but depends only linearly on the masses. We note that our model enables inferring all these physical parameters without resorting to theoretical stellar models, and their accuracy depends, therefore, only on the validity of the model assumptions.

Concerning the orbital eccentricities, \citet{becker2015} obtained upper limits using long-term dynamical simulations and concluded that the orbits would be unstable for eccentricities above 0.06. This result was later used as a prior for their TTV analysis. \citet{dai2015} assumed circular orbits for the three interior planets. Here, we obtain stringent eccentricity upper limits directly from the analysis of the data, without requiring additional assumptions about the long-term stability of the system. The eccentricity 99\% upper confidence limit for planets~e, b, and d are 0.17, 0.014, and 0.033, respectively. 

The dynamical interactions between the planets allow constraining the true mass of the non-transiting planet~c. The inclination is constrained to be higher than 18\degree\ at 99\%, with a corresponding mass upper limit of 4.7~\Mjup. On the other hand, the longitude of the ascending node for this planet is completely unconstrained. The model permits even retrograde orbits, which would not seem to produce a detectable effect on the timescale of the observations. Long-term stability arguments should probably distinguish between prograde and retrograde orbits.

The limb-darkening parameters are precisely constrained thanks to the presence of a giant transiting planet, which produces high signal-to-noise ratio transits. This produces precisely determined radius ratios, which are usually covariate with the limb-darkening coefficients. This is particularly important for the smaller planets in the system, whose transits do not provide strong information on the stellar flux distribution across the disk.

\begin{figure*}[!ht]
  \hspace{-0.2cm}\includegraphics[width=6.3cm]{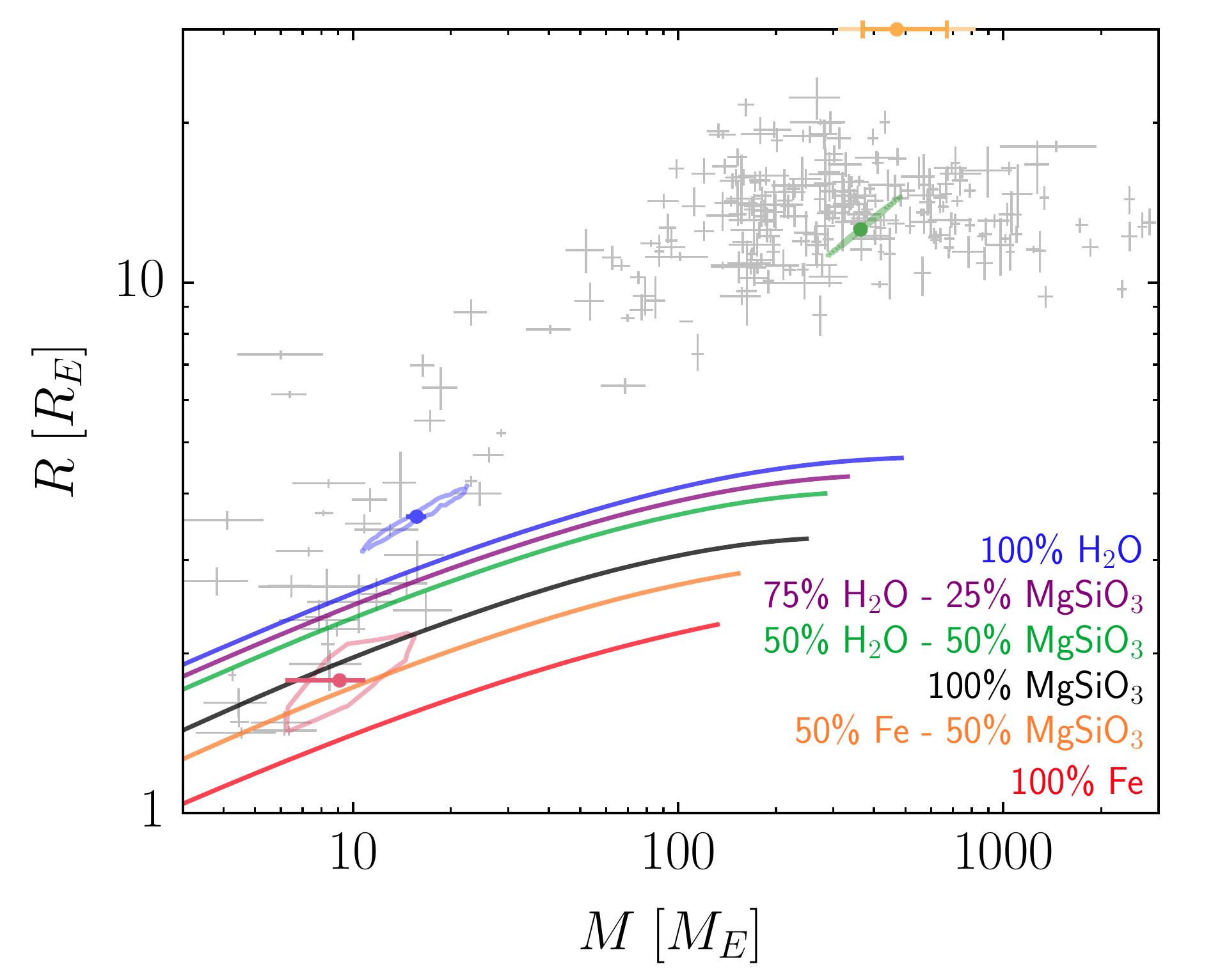}\includegraphics[width=6.3cm]{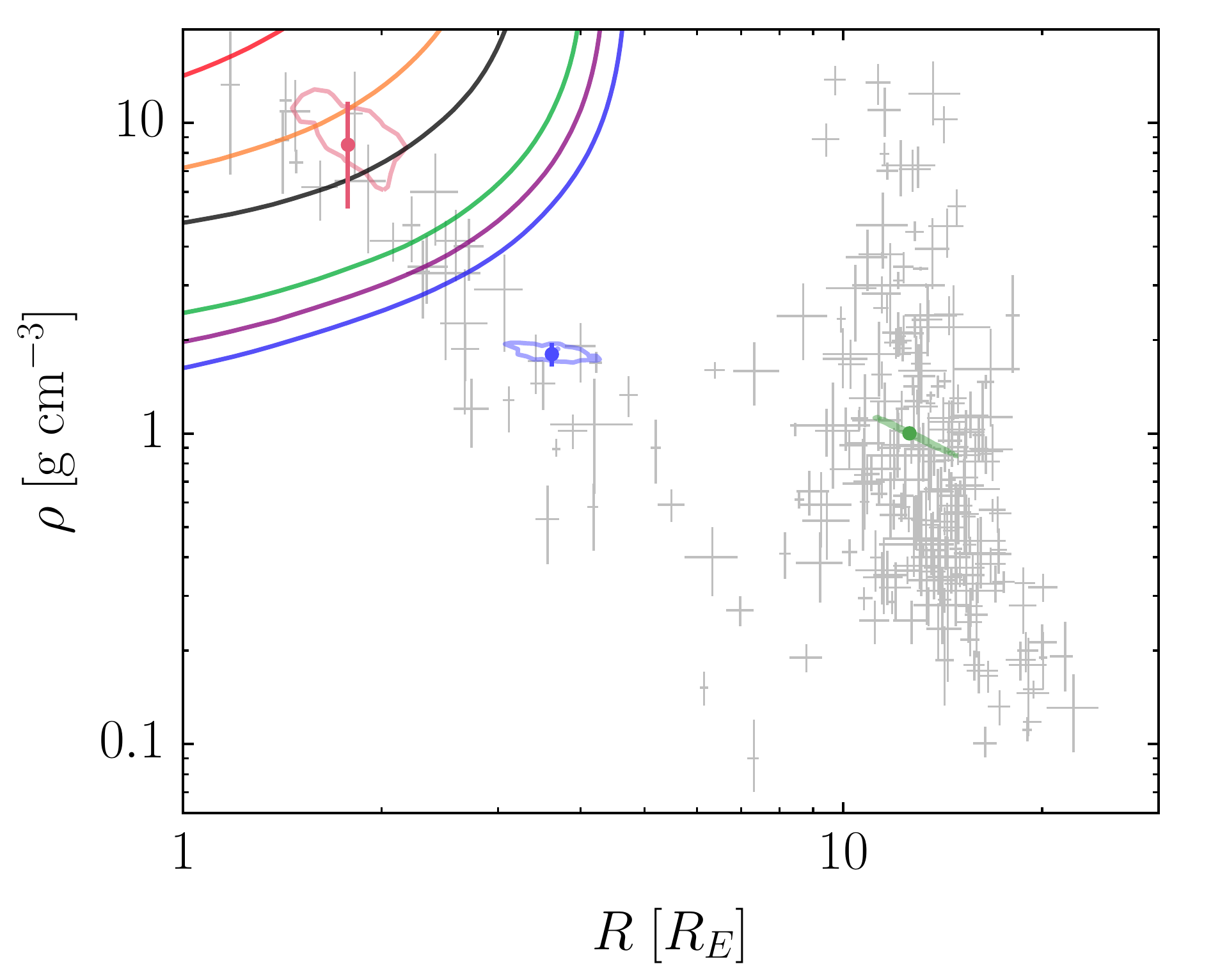}\includegraphics[width=6.3cm]{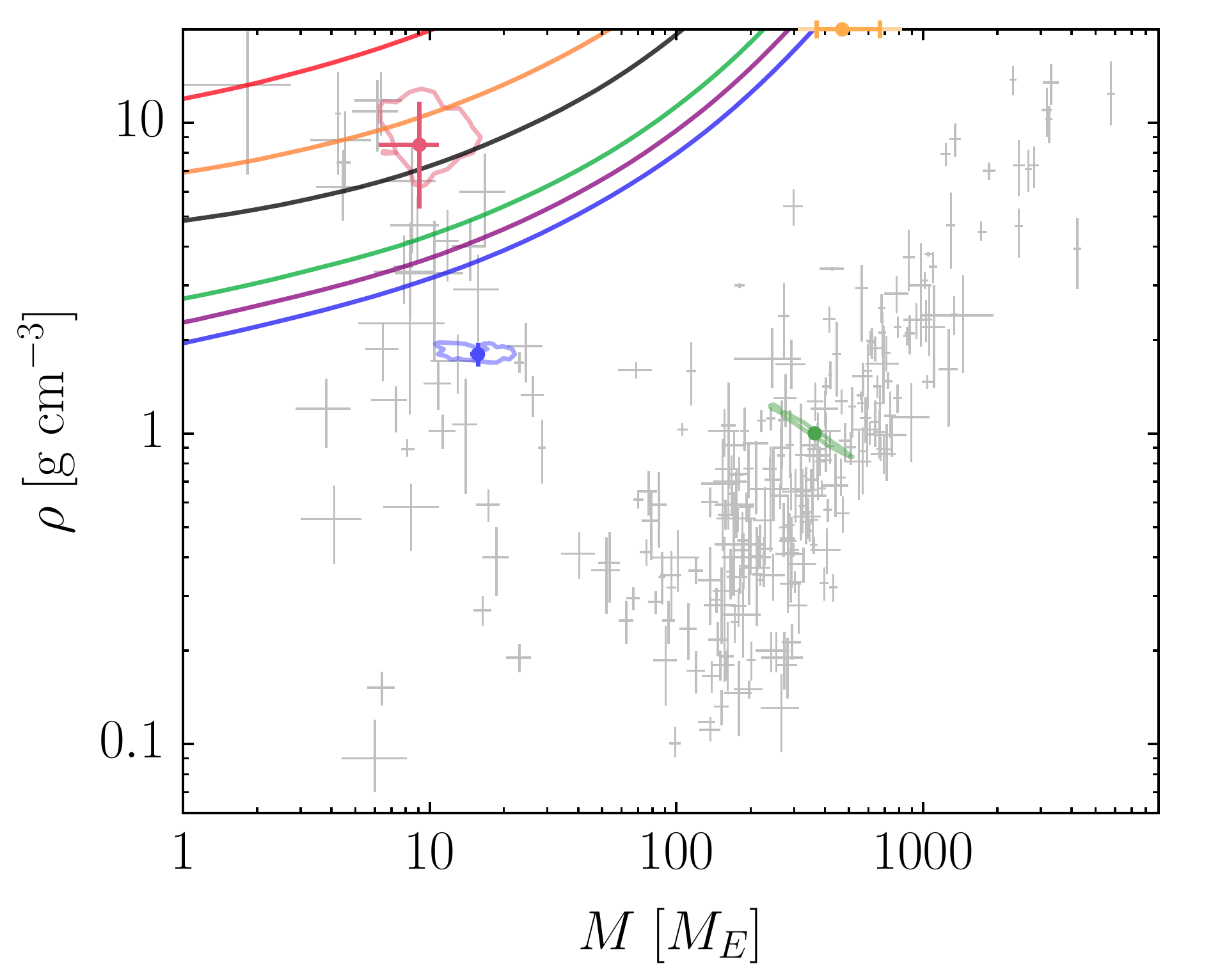}
    \caption{Mass-radius, radius-density, and mass-density diagrams for known exoplanets with parameters determined with a precision $<$50\% \citep{han2014}. The WASP-47 planets are shown as 68.3\% coloured contours for the model-free photodynamical analysis and as dots with 1-$\sigma$ error bars for the analysis using the Dartmouth model (planet~b in green, planet~c in orange, planet~d in blue, and planet~e in red). Planet~c, whose radius and density are not known, is represented at the top of the mass-radius and mass-density diagrams. The solid lines represent theoretical models for different composition \citep{zeng2013}.}
    \label{fig.MR}
\end{figure*}

\subsection{Stellar models}
To improve on the precision of our results, at the expense of accuracy, we added the constraints coming from the theoretical stellar models. To do this, we kept only the posterior samples that are compatible with the Dartmouth models \citep{dotter2008}, given the bulk density from our analysis and the stellar atmospheric parameters from \citet{mortier2013}: T$_{\rm eff}$ = 5576$\pm$68~K, [Fe/H] = 0.36$\pm$0.05. This significantly narrowed the posterior distributions of the stellar mass and radius, and, therefore, those of the planetary parameters. Curiously, the MAP estimate of the stellar mass agrees with the value obtained using stellar models. In Table~\ref{table.results} we report in the last column the parameters for which the precision was improved significantly with respect to our model-free analysis.

In addition to improving the precision of the masses, radii, and semi-major axes for all the planets, the inclusion of theoretical stellar models also narrowed the posterior distributions of other parameters. Radial velocities set stringent constraints on the properties of planet~b, but are much less efficient in determining the mass of planet~d, for which the dynamical interaction with the remaining planets is a major asset. As a consequence, including stellar models improved the determination of the radial velocity amplitude induced by planet~d, which depends on the ratio $M_d/M_\star^{2/3}$, but did not affect the determination of the mass ratio $M_d/M_\star$ significantly. Conversely, the mass ratio $M_b/M_\star$ was improved by the inclusion of the models but not the radial velocity amplitude. As a corollary, given that the mean stellar density and the radius ratios were not affected by the inclusion of the theoretical models, the bulk density of planet~b and the surface gravity of planet~d were improved as well\footnote{The bulk density is $\rho_p~=~\rho_\star\left(M_p/M_\star\right)\left(R_p/R_\star\right)^{-3}$, while the surface gravity is $g_p~=~\mathcal G~M_p/R_p^2~\propto~\rho_\star^{2/3}~M_p/M_\star^{2/3}\left(R_p/R_\star\right)^{-2}$.}. This allows distinguishing whether the mass detection for a given planet is dominated by the radial velocity or timing measurements.  

Concerning the stellar age, the Dartmouth models provided an age of of 7.1$\pm$1.5~Gyr. Using the projected velocity $\vsini = 1.8^{+0.24}_{-0.16}$ \kms\ from \citet{sanchis2015} and the radius from the Dartmouth tracks and assuming the stellar rotational axis lies in the plane of the sky, $i_\star=90$\degree, we derived a rotational period of P$_{\rm rot}=31.2\pm3.5$~days. Coupling this with the mass determination, we derived a gyrochronological age of 6.5$^{+2.6}_{-1.2}$~Gyr \citep{barnes2010,barneskim2010,meibom2015}. Both age determinations agree well, but they disagree with a previous gyrochronological estimation by \citet{brown2014} and \citet{hellier2012}, who used a slightly higher value of $\vsini=3.0\pm0.6$.

The next three transit windows of planet~c are $2\;457\;741\pm22$, $2\;458\;327\pm30$, and $2\;458\;914\pm35$ BJD. The expected centre transit duration for a Jupiter-sized planet is 14.0$\pm$1.6~h. If the transit windows are reduced with follow-up radial velocity observations, this could be a target for the upcoming CHEOPS mission \citep{broeg2013}.

\section{Discussion}
Our analysis allowed characterising the planets and the star of the WASP-47 system without resorting to theoretical stellar models at all, except for the implicit assumption of spherical bodies. The self-consistent modelling is mandatory to optimally exploit the data of this system and led to an improved determination for the planetary bulk densities for some of the planets with respect to previous studies. In addition, these measurements are probably more accurate as well. 

In Fig.~\ref{fig.MR} we placed the WASP-47 planets in mass-radius, mass-density, and radius-density diagrams, together with the planets reported in the literature \citep{han2014}. Our analysis confirms that the structure of planet~d must be dominated by volatiles, and the rocky character of planet~e. Although \citet{rogers2015} determined that most planets with radii larger that 1.6 \Rearth\ are not dense enough to be rocky, the model predicts a non-zero fraction of rocky planets as large as WASP-47~e.

The habitable zone (HZ) of WASP-47 extends between 1.06 au and 1.8 au \citep{kopparapu2013}. The MAP estimate of the semi-major axis of the orbit of WASP-47 c is 1.37 au, but owing to the orbital eccentricity, the planet makes excursions outside of the HZ (Fig.~\ref{fig.HZ}). However, \citet{williamspollard2002} argued that long-term climate stability is dictated by the mean incident flux throughout the orbit rather than by the time spent in the HZ. The effective incident flux on WASP-47 c is $0.640 \pm 0.052$ times the flux received by Earth, which would place it in the middle of the HZ \citep[see][fig. 8]{diaz2016b, kopparapu2013}. Therefore, hypothetical rocky satellites orbiting WASP-47 c have good prospects for being habitable. 

Including theoretical stellar models improves the precision for some parameters, in particular masses and radii, but does not change our conclusions about the planet compositions significantly. This is probably because WASP-47 is a solar-type star for which models are expected to perform correctly. On the other hand, these determinations are probably less accurate, and might exhibit systematic errors larger than the measured precision.

The current precision of the photometric measurements of extrasolar planets already calls for the use of realistic models including the interactions between the planets. In the coming years, TESS \citep{ricker2014} and PLATO \citep{rauer2014} will continue to provide such high-quality light curves, but the brightness of the typical target will also permit obtaining high-precision radial velocity measurements, further increasing the need for realistic self-consistent modelling of the data.


\begin{acknowledgements}
  We thank C. Hellier, who kindly sent us the EulerCam transit data, R. Mardling for discussions about dynamics, T. Fenouillet and Y. Revaz for their assistance with the computing clusters used in this work, and L. Kreidberg for her Mandel \& Agol code. This paper includes data collected by the Kepler mission. Funding for the Kepler mission is provided by the NASA Science Mission directorate. This research has made use of the Exoplanet Orbit Database and the Exoplanet Data Explorer at exoplanets.org. Simulations in this paper made use of the \reb\ code which can be downloaded freely at \texttt{http://github.com/hannorein/rebound}. Part of these simulations have been run on the {\it Regor} cluster kindly provided by the Observatoire de Gen\`eve. JMA and XB acknowledges funding from the European Research Council under the ERC Grant Agreement n. 337591-ExTrA. This work has been carried out within the frame of the National Centre for Competence in Research PlanetS supported by the Swiss National Science Foundation (SNSF).
\end{acknowledgements}

\bibliographystyle{aa}
\bibliography{WASP-47}

\begin{appendix}
\section{Additional figures and tables}
\begin{table*}
  \tiny
\renewcommand{\arraystretch}{1.25}
\centering
\caption{Inferred system parameters: MAP, 95\% HDI, posterior median, and 68.3\% CI for the model-free photodynamical analysis, and the latter for the precision-improved parameters using theoretical stellar models. The astrocentric orbital elements are given for the reference time $t_{\mathrm{ref}} = 2\;456\;979.5$~BJD$_{\mathrm{TDB}}$.}\label{table.results}
\begin{tabular}{lccccc}
\hline
Parameter & & MAP &  95\% HDI & Median  &Stellar models \\
&  & & & and 68.3\% CI & median and 68.3\% CI \\
\hline

\emph{\bf Star} \smallskip\\

Stellar mass, $M_\star$                             & [\Msun]         & 1.0174  & [0.242, 3.126] & 1.11$^{+0.89}_{-0.49}$ & 1.029$\pm$0.031 \\
Stellar radius, $R_\star$$^{\bullet}$               & [\Rnom]         & 1.1299  & [0.804, 1.698] & 1.16$\pm$0.26          & 1.132$\pm$0.012 \\
Stellar mean density, $\rho_{\star}$$^{\bullet}$    & [$\mathrm{g\;cm^{-3}}$]   & 0.9943  & [0.9713, 1.0259] & 0.999$\pm$0.015     & \\
Surface gravity, \logg\                             & [cgs]           & 4.33946 & [4.2069, 4.5192] & 4.354$\pm$0.084      & 4.3428$\pm$0.0058 \smallskip\\

$q_1$$^{\dagger,\bullet}$ K2                        &                 & 0.3864 & [0.3585, 0.4179] & 0.387$\pm$0.015     & \\
$q_2$$^{\dagger,\bullet}$ K2                        &                 & 0.4279 & [0.3984, 0.4642] & 0.430$\pm$0.016     & \\
Linear limb darkening, $u_{\mathrm{a}}$ K2          &                 & 0.5319 & [0.51646, 0.55534] & 0.535$\pm$0.010   & \\
Quadratic darkening, $u_{\mathrm{b}}$ K2            &                 & 0.0897 & [0.0455, 0.1333] & 0.087$\pm$0.021    & \\

$q_1$$^{\dagger,\bullet}$ EulerCam                  &                 & 0.2929 & [0.1912, 0.3903] & 0.282$\pm$0.052     & \\
$q_2$$^{\dagger,\bullet}$ EulerCam                  &                 & 0.643  & [0.427, 0.885]   & 0.64$\pm$0.13       & \\
$u_{\mathrm{a}}$ EulerCam                           &                 & 0.6957 & [0.5398, 0.8123] & 0.682$\pm$0.067     & \\
$u_{\mathrm{b}}$ EulerCam                           &                 & -0.154 & [-0.383, 0.067]  & -0.15$\pm$0.11      & \medskip\\

\emph{\bf Planet b} \smallskip\\

Semi-major axis, $a$                                  & [au]                   & 0.050918     & [0.03631, 0.07665] & 0.052$\pm$0.011              & 0.05111$\pm$0.00051 \\
Eccentricity, $e$                                     &                        & 0.00380      & [0.00000, 0.01052] & 0.0028$^{+0.0042}_{-0.0020}$ & \\
Inclination, $i$$^{\bullet}$                          & [\degree]              & 88.927       & [88.734, 89.403]   & 89.02$\pm$0.17               & \\
Argument of pericentre, $\omega$                      & [\degree]              & 3.747        & [-131.20, 171.20]  & 30$\pm$79                    & \\
Longitude of the ascending node, $\Omega$             & [\degree]              & 180$^{\ast}$ &  &  & \\
Mean anomaly, $M_0$                                   & [\degree]              & 144.870      & [-16.80, 285.60]   & 118$\pm$78                   & \smallskip\\

$\sqrt{e}\cos{\omega}$$^{\bullet}$                    &                        & 0.0615      & [-0.0439, 0.0946]    & 0.028$\pm$0.037            &  \\
$\sqrt{e}\sin{\omega}$$^{\bullet}$                    &                        & 0.0040      & [-0.0543, 0.0916]    & 0.018$\pm$0.042            &  \\
Radius ratio, $R_{\mathrm{p}}/R_\star$$^{\bullet}$    &                        & 0.102036    & [0.101564, 0.102303] & 0.10193$\pm$0.00018        & \\
Mass ratio, $M_{\mathrm{p}}/M_\star$$^{\bullet}$      &                        & 0.0010609    & [0.000667, 0.001416] & 0.00104$\pm$0.00024        & 0.001061$\pm$1.7$\e{-5}$ \\
Scaled semi-major axis, $a/R_{\star}$                 &                        & 9.6901      & [9.6125, 9.7899]     & 9.705$\pm$0.047            &   \\
$T_0'$$^{\bullet}$\;-\;2\;450\;000                    & [BJD$_{\mathrm{TDB}}$] & 6982.978187 & [6982.977745, 6982.978868] & 6982.97823$\pm$0.00032 & 6982.97826$\pm$0.00016 \\
$P'$$^{\bullet}$                                      & [d]                    & 4.160666    & [4.160061, 4.161365] & 4.16071$\pm$0.00038        & 4.16075$\pm$0.00016 \\
$K'$                                                  & [\ms]                  & 141.22      & [138.43, 145.19]     &  142.0$\pm$1.7             &  \smallskip\\

Planet mass, $M_{\mathrm{p}}$                         &[\Mearth]               & 359.37  & [174.61, 769.82]   & 383$^{+190}_{-120}$    & 363.8$\pm$8.6 \\
                                                      &[\Mjup]                 & 1.1307  & [0.549, 2.422]     & 1.21$^{+0.59}_{-0.39}$ & 1.145$\pm$0.027 \\
Planet radius, $R_{\mathrm{p}}$                       &[\Renom]                & 12.576  & [8.95, 18.91]      & 12.9$\pm$2.8           & 12.59$\pm$0.14 \\
                                                      &[\RJnom]                 & 1.1219  & [0.798, 1.687]     & 1.15$\pm$0.25          & 1.123$\pm$0.013 \\
Planet mean density, $\rho_{\mathrm{p}}$              &[$\mathrm{g\;cm^{-3}}$] & 0.9929  & [0.627, 1.376]     & 0.98$\pm$0.21          & 1.001$\pm$0.023 \\
Planet surface gravity, $\log$\,$g_{\mathrm{p}}$      &[cgs]                   & 3.34763 & [3.33826, 3.36825] & 3.3522$\pm$0.0071      & \medskip\\

\emph{\bf Planet c} \smallskip\\

Semi-major axis, $a$                                  & [au]                   & 1.3694  & [0.973, 2.060]   & 1.41$\pm$0.30  & 1.375$\pm$0.019 \\
Eccentricity, $e$                                     &                        & 0.366   & [0.115, 0.575]   & 0.36$\pm$0.12  &   \\
Inclination, $i$$^{\bullet}$                          & [\degree]              & 72.829  & [21.93, 151.28]  & 87$\pm$45      &  \\
Argument of pericentre, $\omega$                      & [\degree]              & 141.946 & [72.04, 172.82]  & 136$\pm$20     &  \\
Longitude of the ascending node, $\Omega$$^{\bullet}$ & [\degree]              & 167.77  & [0, 360]         & [0, 360]       & \\
Mean anomaly, $M_0$                                   & [\degree]              & 222.333 & [201.60, 288.00] & 229$\pm$15     &   \smallskip\\

$\sqrt{e}\cos{\omega}$$^{\bullet}$                    &                        & -0.4767  & [-0.669, 0.052]      & -0.41$^{+0.19}_{-0.13}$          & \\
$\sqrt{e}\sin{\omega}$$^{\bullet}$                    &                        & 0.373    & [0.063, 0.645]       & 0.40$\pm$0.14                    & \\
Mass ratio, $M_{\mathrm{p}}/M_\star$$^{\bullet}$      &                        & 0.001375 & [0.000565, 0.002547] & 0.00133$^{+0.00062}_{-0.00038}$  & \\
Scaled semi-major axis, $a/R_{\star}$                 &                        & 260.60   & [258.94, 263.72]     & 261.4$\pm$1.3                    &  \\
$T_0'$$^{\bullet}$\;-\;2\;450\;000                    & [BJD$_{\mathrm{TDB}}$] & 7162.464 & [7118.25, 7182.95]   & 7156$\pm$18                      & \\
$P'$$^{\bullet}$                                      & [d]                    & 580.289  & [561.83, 594.05]     & 580.7$\pm$9.6                    & \\
$K'$                                                  & [\ms]                  & 36.241   & [18.84, 43.52]       & 29.4$\pm$6.1                     &      \smallskip\\

Planet mass, $M_{\mathrm{p}}$                         &[\Mearth]               & 465.688 & [101.22, 1116.23] & 500$^{+320}_{-190}$       & 470$^{+200}_{-100}$ \\
                                                      &[\Mjup]                 & 1.465   & [0.318, 3.512]    & 1.57$^{+1.0}_{-0.59}$     & 1.47$^{+0.64}_{-0.33}$ \\
$M_{\mathrm{p}}\sin{i}$                               &[\Mearth]               & 444.932 & [150.95, 845.06]  & 380$^{+220}_{-130}$       & 361$^{+80}_{-54}$ \\
                                                      &[\Mjup]                 & 1.3999  & [0.475, 2.659]    & 1.21$^{+0.69}_{-0.41}$    & 1.13$^{+0.25}_{-0.17}$ \smallskip\\

\hline
\end{tabular}
\end{table*}

\begin{table*}
  \tiny
\renewcommand{\arraystretch}{1.25}
\centering
\caption{Continuation Table~\ref{table.results}}
\begin{tabular}{lccccc}
\hline
Parameter & & MAP &  95\% HDI & Median  &Stellar models \\
&  & & & and 68.3\% CI & median and 68.3\% CI \\
\hline
\emph{\bf Planet d} \smallskip\\

Semi-major axis, $a$                                  & [au]                   & 0.085769 & [0.0613, 0.1289]              & 0.088$\pm$0.019              & 0.08609$\pm$0.00087 \\
Eccentricity, $e$                                     &                        & 0.00752  & [0.00000, 0.02388]            & 0.0060$^{+0.0098}_{-0.0041}$ & \\
Inclination, $i$$^{\bullet}$                          & [\degree]              & 90.839   & [88.965, 89.515]$^{\ddagger}$ & 89.22$\pm$0.13$^{\ddagger}$  & \\
Argument of pericentre, $\omega$                      & [\degree]              & 4.655    & [-116.80, 185.60]             & 27$\pm$89                    & \\
Longitude of the ascending node, $\Omega$$^{\bullet}$ & [\degree]              & 178.736  & [174.12, 188.44]              & 181.4$\pm$4.1                & \\
Mean anomaly, $M_0$                                   & [\degree]              & 93.201   & [-81.20, 221.19]              & 68$\pm$87                    & \smallskip\\

$\sqrt{e}\cos{\omega}$$^{\bullet}$                    &                        & 0.0865        & [-0.0683, 0.1511]      & 0.033$\pm$0.065         & \\
$\sqrt{e}\sin{\omega}$$^{\bullet}$                    &                        & 0.0070        & [-0.0987, 0.1331]      & 0.020$\pm$0.066         & \\
Radius ratio, $R_{\mathrm{p}}/R_\star$$^{\bullet}$    &                        & 0.029264      & [0.029004, 0.029622]   & 0.02931$\pm$0.00015     & \\
Mass ratio, $M_{\mathrm{p}}/M_\star$$^{\bullet}$      &                        & 4.564$\e{-5}$ & [3.792, 5.280]$\e{-5}$ & (4.54$\pm$0.38)$\e{-5}$ & \\
Scaled semi-major axis, $a/R_{\star}$                 &                        & 16.3223       & [16.1715, 16.4699]     & 16.327$\pm$0.079        & \\
$T_0'$$^{\bullet}$\;-\;2\;450\;000                    & [BJD$_{\mathrm{TDB}}$] & 6988.37565    & [6988.3527, 6988.3993] & 6988.375$\pm$0.014      & 6988.3770$\pm$0.0015 \\
$P'$$^{\bullet}$                                      & [d]                    & 9.09585       & [9.0740, 9.1215]       & 9.095$\pm$0.014         & 9.0961$\pm$0.0012 \\
$K'$                                                  & [\ms]                  & 4.681         & [3.434, 6.300]         & 4.77$\pm$0.81           & 4.71$\pm$0.31 \smallskip\\

Planet mass, $M_{\mathrm{p}}$                         &[\Mearth]               & 15.46  & [3.99, 44.34]                 & 16.8$^{+12}_{-7.0}$          & 15.7$\pm$1.1 \\
Planet radius, $R_{\mathrm{p}}$                       &[\Renom]                & 3.6067 & [2.565, 5.470]                & 3.71$\pm$0.82                & 3.619$\pm$0.044 \\
Planet mean density, $\rho_{\mathrm{p}}$              &[$\mathrm{g\;cm^{-3}}$] & 1.811  & [1.490, 2.097]                & 1.80$\pm$0.16                &  \\
Planet surface gravity, $\log$\,$g_{\mathrm{p}}$      &[cgs]                   & 3.0661 & [2.9480, 3.1967]              & 3.074$\pm$0.067              & 3.068$\pm$0.030 \medskip\\

\emph{\bf Planet e} \smallskip\\

Semi-major axis, $a$                                  & [au]                   & 0.016816 & [0.01199, 0.02532]         & 0.0173$\pm$0.0038         & 0.01688$\pm$0.00017 \\
Eccentricity, $e$                                     &                        & 0.0160   & [0.0000, 0.1176]           & 0.030$^{+0.036}_{-0.020}$ & \\
Inclination, $i$$^{\bullet}$                          & [\degree]              & 91.82    & [83.15, 97.56]             & 86.2$\pm$2.1$^{\ddagger}$ & \\
Argument of pericentre, $\omega$                      & [\degree]              & 84.92    & [0, 360]                   & 160$\pm$140               & \\
Longitude of the ascending node, $\Omega$$^{\bullet}$ & [\degree]              & 193.34   & [120.60, 233.53]           & 187$\pm$33                & \\
Mean anomaly, $M_0$                                   & [\degree]              & 244.10   & [0, 360]                   & 190$\pm$120               &  \smallskip\\

$\sqrt{e}\cos{\omega}$$^{\bullet}$                    &                        & 0.011         & [-0.282, 0.306]            & -0.02$\pm$0.18            &  \\
$\sqrt{e}\sin{\omega}$$^{\bullet}$                    &                        & 0.1260        & [-0.2911, 0.1952]          & 0.034$^{+0.088}_{-0.14}$  & \\
Radius ratio, $R_{\mathrm{p}}/R_\star$$^{\bullet}$    &                        & 0.014328      & [0.014066, 0.014774]       & 0.01439$\pm$0.00016       & \\
Mass ratio, $M_{\mathrm{p}}/M_\star$$^{\bullet}$      &                        & 3.026$\e{-5}$ & [0.975, 4.591]$\e{-5}$     & (2.54$\pm$0.93)$\e{-5}$   &  \\
Scaled semi-major axis, $a/R_{\star}$                 &                        & 3.2001        & [3.1748, 3.2334]           & 3.205$\pm$0.015           & \\
$T_0'$$^{\bullet}$\;-\;2\;450\;000                    & [BJD$_{\mathrm{TDB}}$] & 6979.765020   & [6979.763101, 6979.765758] & 6979.76455$\pm$0.00094    & \\
$P'$$^{\bullet}$                                      & [d]                    & 0.7896264     & [0.7895968, 0.7896653]     & 0.789636$\pm$1.7$\e{-5}$  &  \\
$K'$                                                  & [\ms]                  & 7.01          & [2.21, 9.32]               & 6.1$\pm$1.9               & \smallskip\\

Planet mass, $M_{\mathrm{p}}$                         &[\Mearth]               & 10.25  & [1.77, 22.70]                & 9.1$^{+5.5}_{-3.6}$       & 9.1$^{+1.8}_{-2.9}$ \\
Planet radius, $R_{\mathrm{p}}$                       &[\Renom]                & 1.7659 & [1.256, 2.692]               & 1.82$\pm$0.40             & 1.778$\pm$0.031 \\
Planet mean density, $\rho_{\mathrm{p}}$              &[$\mathrm{g\;cm^{-3}}$] & 10.23  & [3.39, 15.98]                & 8.5$\pm$3.2               & 8.8$\pm$2.7  \\
Planet surface gravity, $\log$\,$g_{\mathrm{p}}$      &[cgs]                   & 3.508  & [3.026, 3.695]               & 3.45$^{+0.10}_{-0.16}$    &  \medskip\\

\emph{\bf Data} \smallskip\\

K2 jitter$^{\bullet}$                 & & 0.89868 & [0.89319, 0.90644] & 0.8997$\pm$0.0034 \\
EulerCam jitter$^{\bullet}$           & & 0.8931  & [0.8449, 1.0135]   & 0.934$\pm$0.040 \\
CORALIE 2012 jitter$^{\bullet}$       & & 1.093   & [0.736, 1.597]     & 1.09$\pm$0.25 \\
CORALIE 2016a jitter$^{\bullet}$      & & 0.828   & [0.711, 1.405]     & 1.03$\pm$0.19 \\
CORALIE 2016b jitter$^{\bullet}$      & & 0.372   & [0.216, 1.438]     & 0.57$^{+0.35}_{-0.19}$ \\
PFS jitter$^{\bullet}$                & & 2.242   & [1.605, 3.153]     & 2.31$\pm$0.39 \smallskip\\

CORALIE 2012 offset$^{\bullet}$  & [\kms] & 27.07055 & [27.05995, 27.08424]   & 27.0711$\pm$0.0068 \\
CORALIE 2016a offset$^{\bullet}$ & [\kms] & 27.08882 & [27.07833, 27.09232]   & 27.0852$\pm$0.0037 \\
CORALIE 2016b offset$^{\bullet}$ & [\kms] & 27.05459 & [27.04380, 27.07814]   & 27.0612$\pm$0.0078 \\
PFS offset$^{\bullet}$           & [\kms] & -0.02253 & [-0.02696, 0.00107]    & -0.0112$\pm$0.0076 \smallskip\\

\hline

\end{tabular}
\begin{list}{}{}
\item {\bf{Notes.}}
  $^{(\bullet)}$ \emcee\ jump parameter.
  $^{(\dagger)}$ \citet{kipping2013} parametrisation for the limb-darkening coefficients to consider only physical values.
  $^{(\ast)}$ fixed at $t_{\mathrm{ref}}$.\\
  $^{(\ddagger)}$ reflected with respect to $i = 90\degree$, the supplementary angle is equally probable. \\
  $T'_0 \equiv t_{\mathrm{ref}} - \frac{P'}{2\pi}\left(M_0-E+e\sin{E}\right)$ with $E=2\arctan{\left\{\sqrt{\frac{1-e}{1+e}}\tan{\left[\frac{1}{2}\left(\frac{\pi}{2}-\omega\right)\right]}\right\}}$, $P' \equiv \sqrt{\frac{4\pi^2a^{3}}{\mathcal G M_{\star}}}$, $K' \equiv \frac{M_p \sin{i}}{M_\star^{2/3}\sqrt{1-e^2}}\left(\frac{2 \pi \mathcal G}{P'}\right)^{1/3}$.\\ 
  CODATA 2014: $\mathcal G$ = 6.674$\;$08\ten[-11]~$\rm{m^3\;kg^{-1}\;s^{-2}}$. IAU 2012: \rm{au} = 149$\;$597$\;$870$\;$700~\rm{m}$\;$. IAU 2015: \Rnom = 6.957\ten[8]~\rm{m}, \GMnom = 1.327$\;$124$\;$4\ten[20]~$\rm{m^3\;s^{-2}}$, \Renom = 6.378$\;$1\ten[6]~\rm{m}, \GMenom = 3.986$\;$004\ten[14]~$\rm{m^3\;s^{-2}}$, $\RJnom$ = 7.149$\;$2\ten[7]~\rm{m}, \GMJnom = 1.266$\;$865$\;$3\ten[17]~$\rm{m^3\;s^{-2}}$. \\
  $\Msun$ = \GMnom/$\mathcal G$, \Mearth = \GMenom/$\mathcal G$, \Mjup = \GMJnom/$\mathcal G$, $k^2$ = \GMnom$\;(86\;400~\rm{s})^2$/$\rm{au}^3$
\end{list}
\end{table*}

\begin{figure*}[!ht]
    \centering
    \begin{minipage}{0.55\textwidth}
        \centering
        \includegraphics[width=0.99\linewidth]{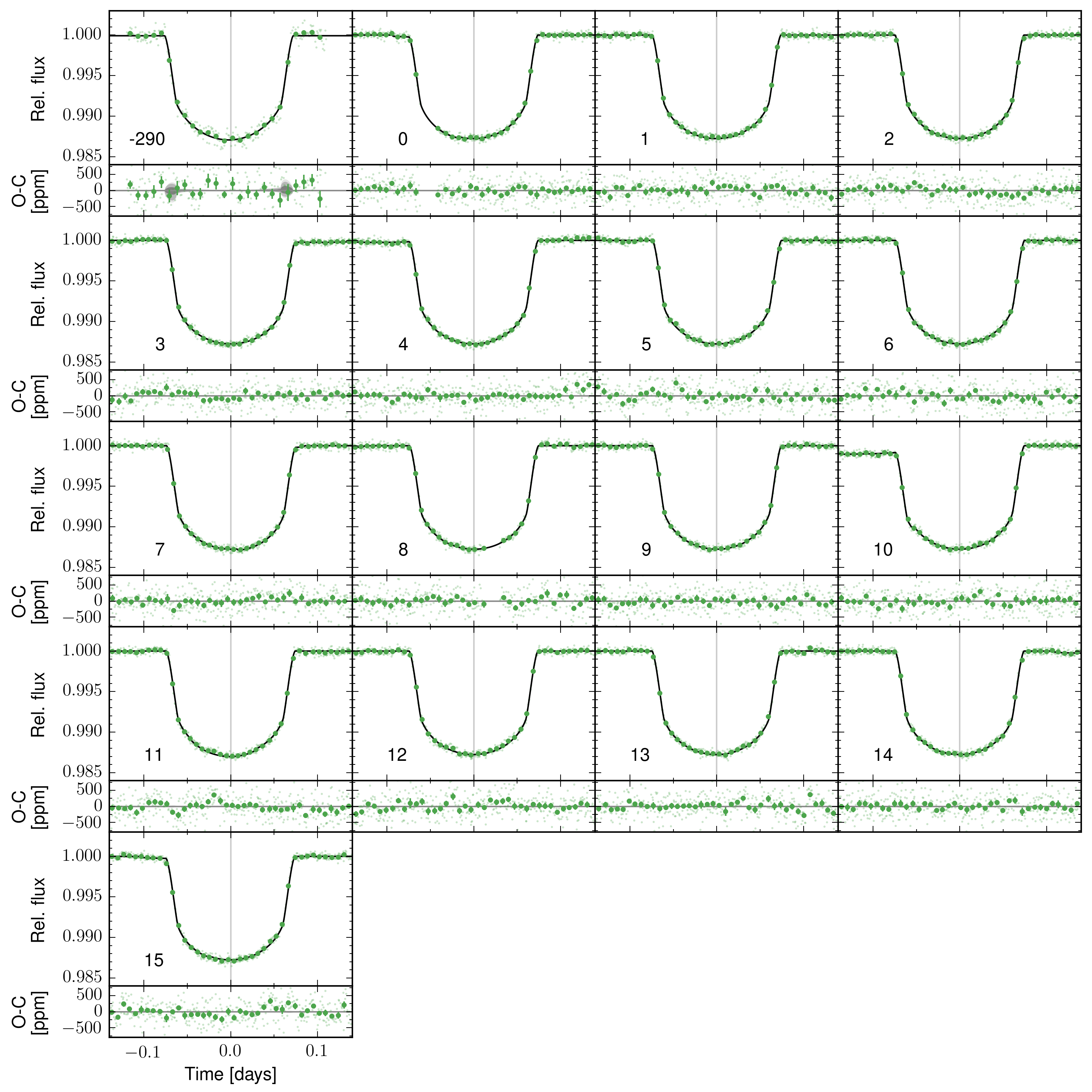}
    \end{minipage}%
    \hspace{0.2cm}
    \begin{minipage}{0.43\textwidth}
        \centering
        \vspace{-2cm}
        \includegraphics[width=0.99\linewidth]{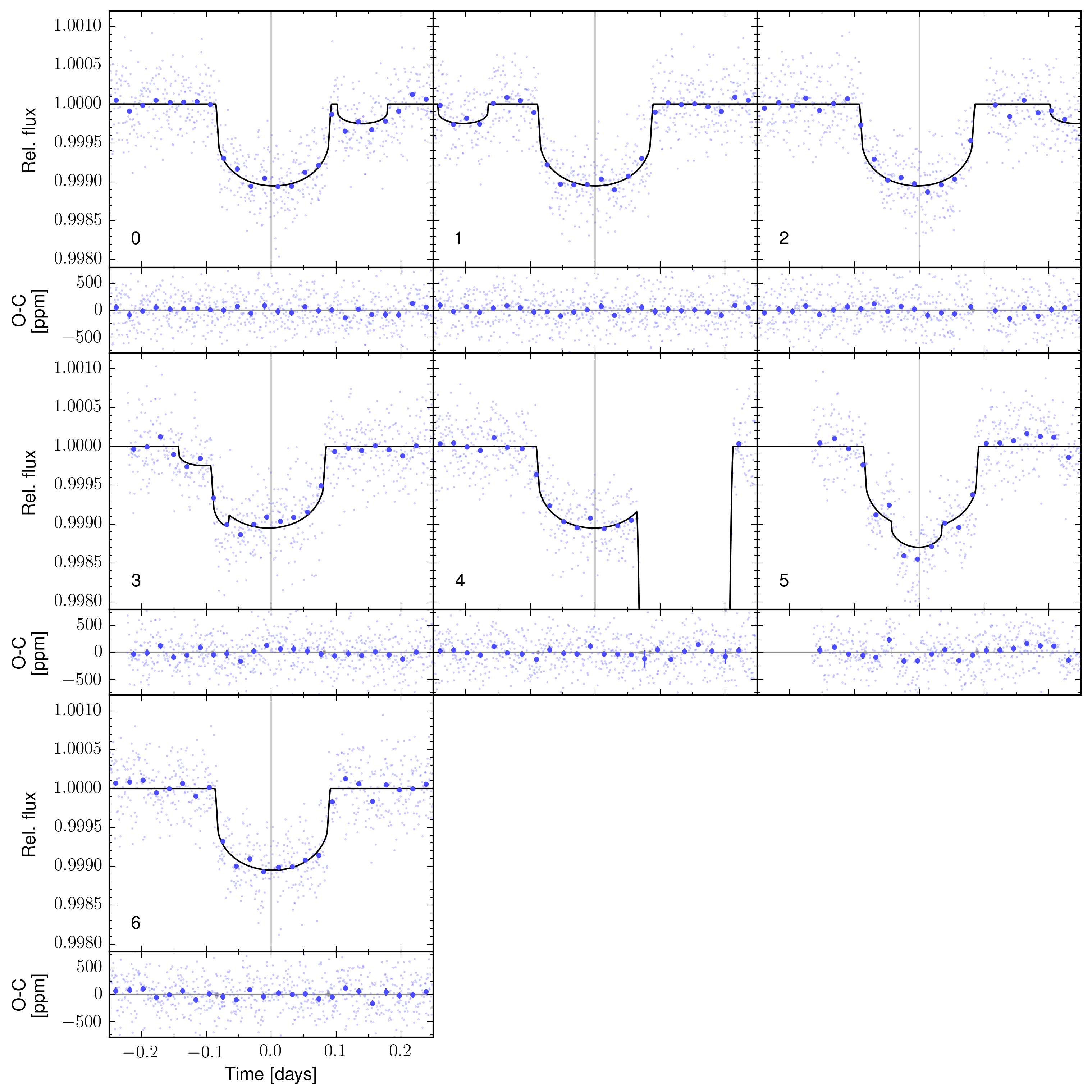}
    \end{minipage}
    \vspace{-1.8cm}
    \\\hspace{3.5cm}\includegraphics[height=14.9cm]{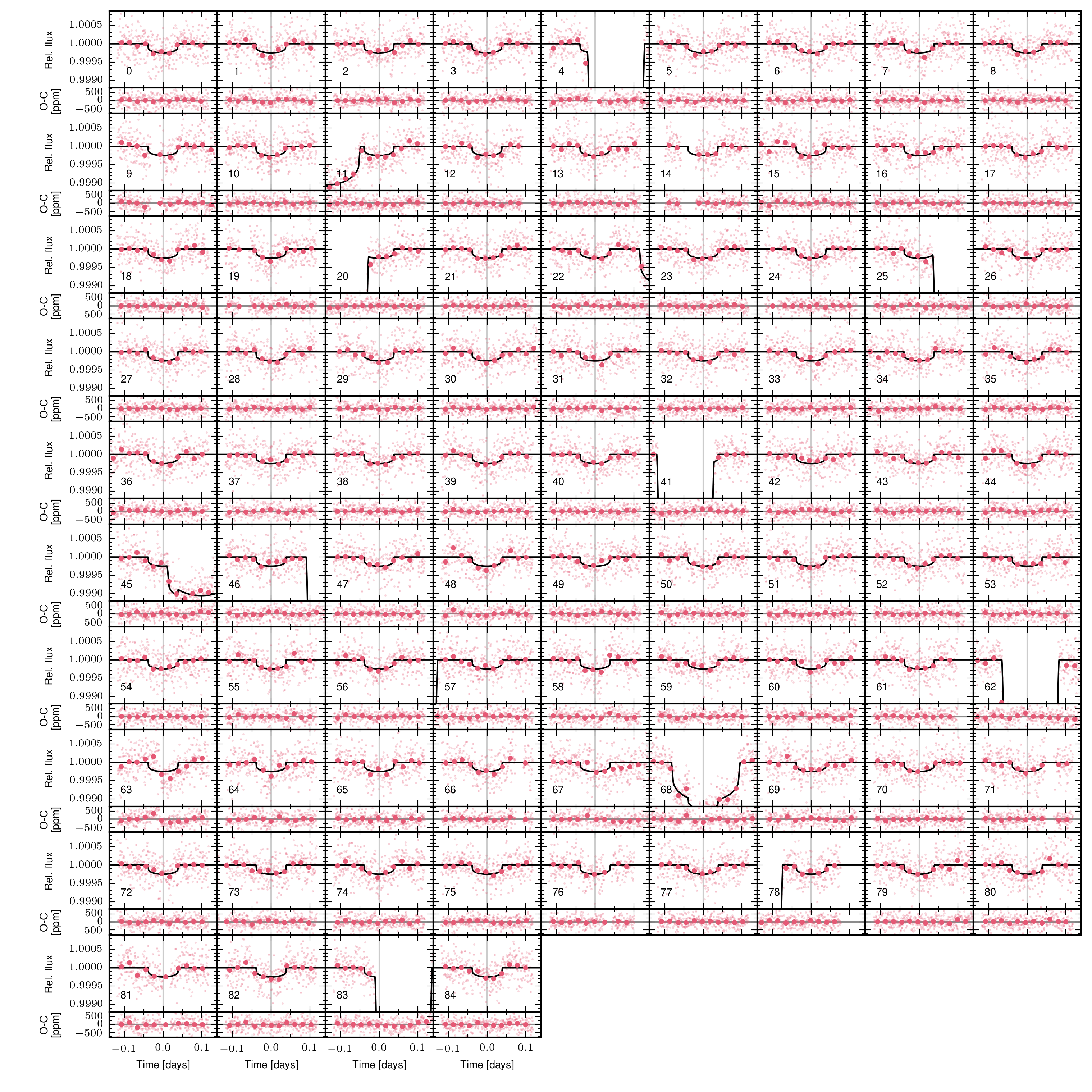}
        \caption{Photometric data of planets~b (top, left), d (top, right), and e (bottom). Each transit is centred relative to a linear ephemeris. Small dots represents the photometric data, whereas circles represent the data binned. Each panel is labelled with the epoch, with zero the first transit after $t_{\mathrm{ref}}$. The black curve is the median value of the distribution of oversampled models corresponding to 1000 random MCMC steps, the different shades of grey represent the 68.3, 95.5, and 99.7\% confidence intervals. In the lower part of each panel the residuals after subtracting the MAP model to the observed data are shown.}
        \label{fig.PH}
\end{figure*}

\begin{figure*}
  \centering
  \includegraphics[height=5.9cm]{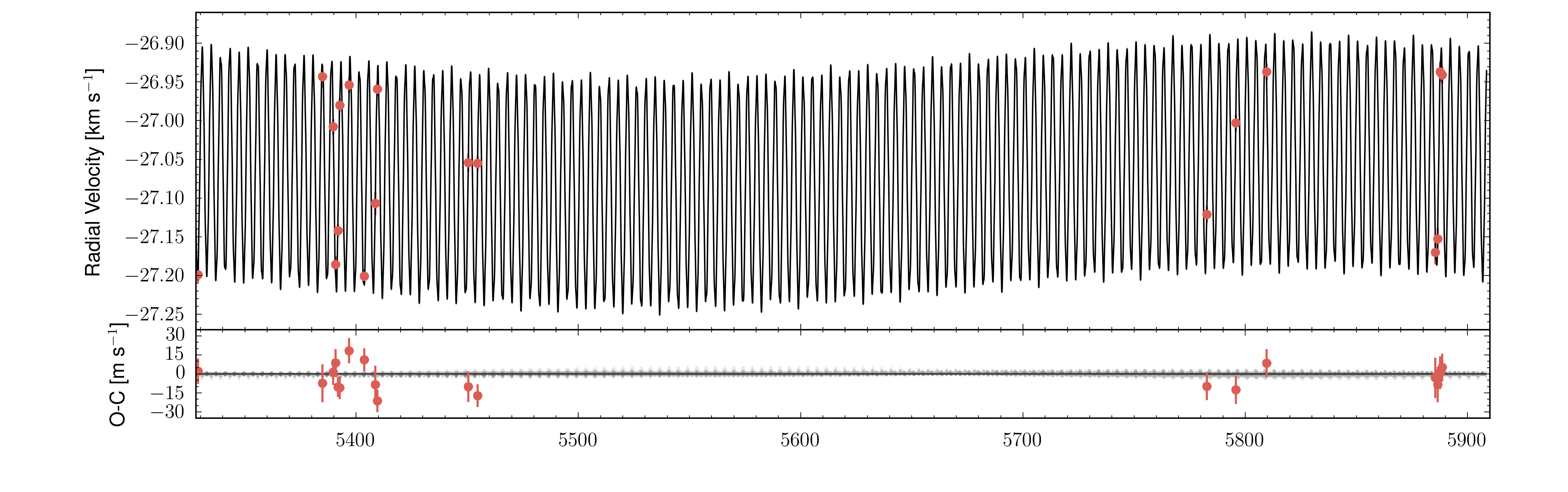}
  
  \includegraphics[height=5.9cm]{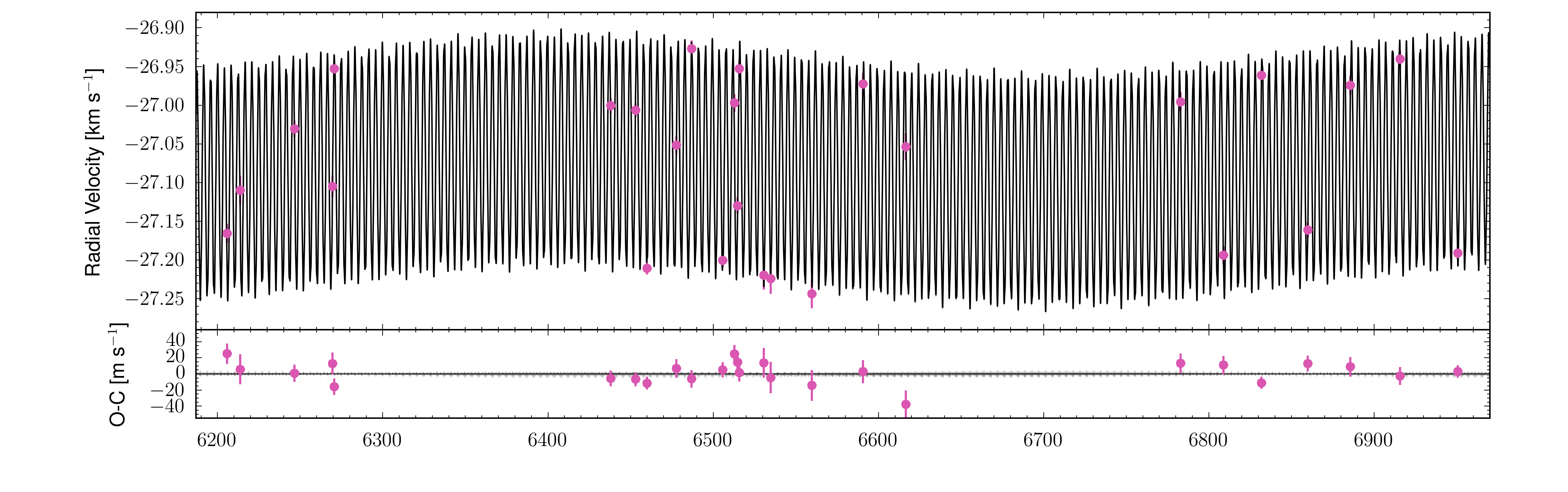}
  
  \includegraphics[height=5.9cm]{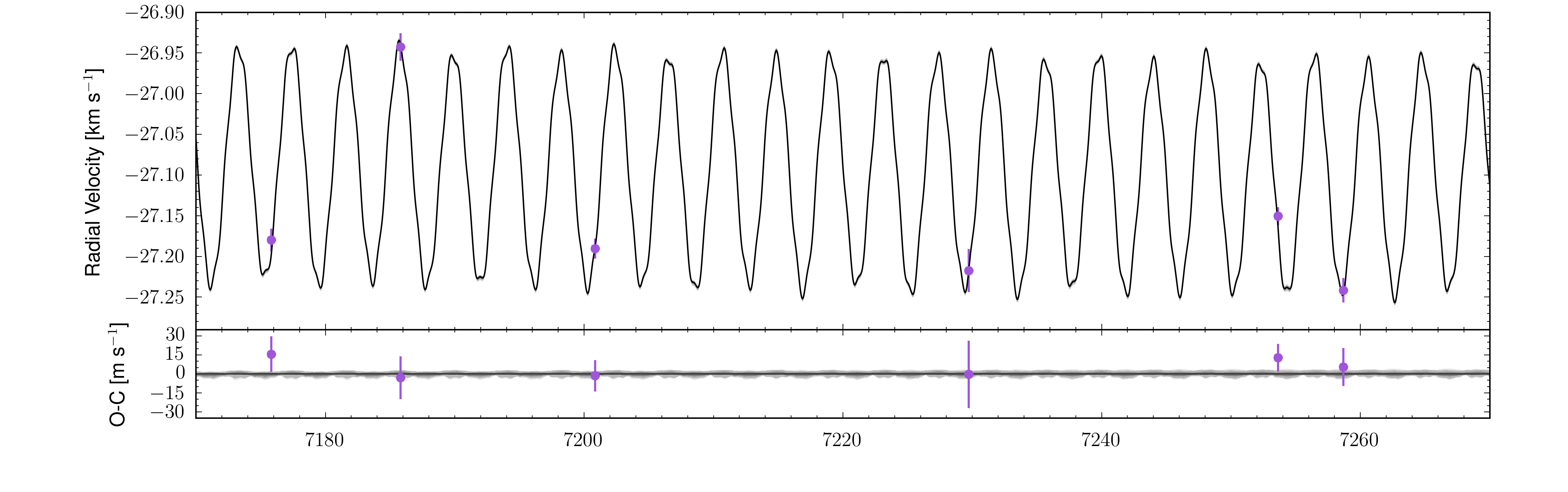}
  
  \includegraphics[height=5.9cm]{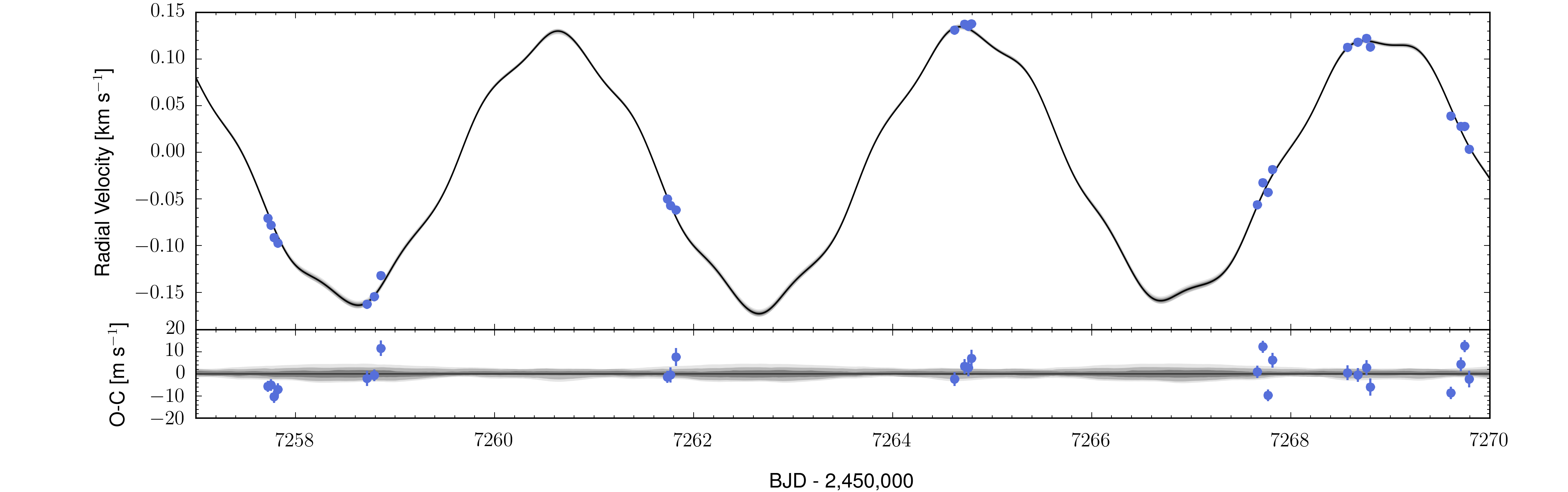}
  \caption{Idem Fig.~\ref{fig.PH} for the radial velocity as a function of time. From top to bottom: CORALIE 2012, CORALIE 2016a, CORALIE 2016b, and PFS.}
  \label{fig.RVt}
\end{figure*}

\begin{figure*}
  \centering
  \includegraphics[height=14cm]{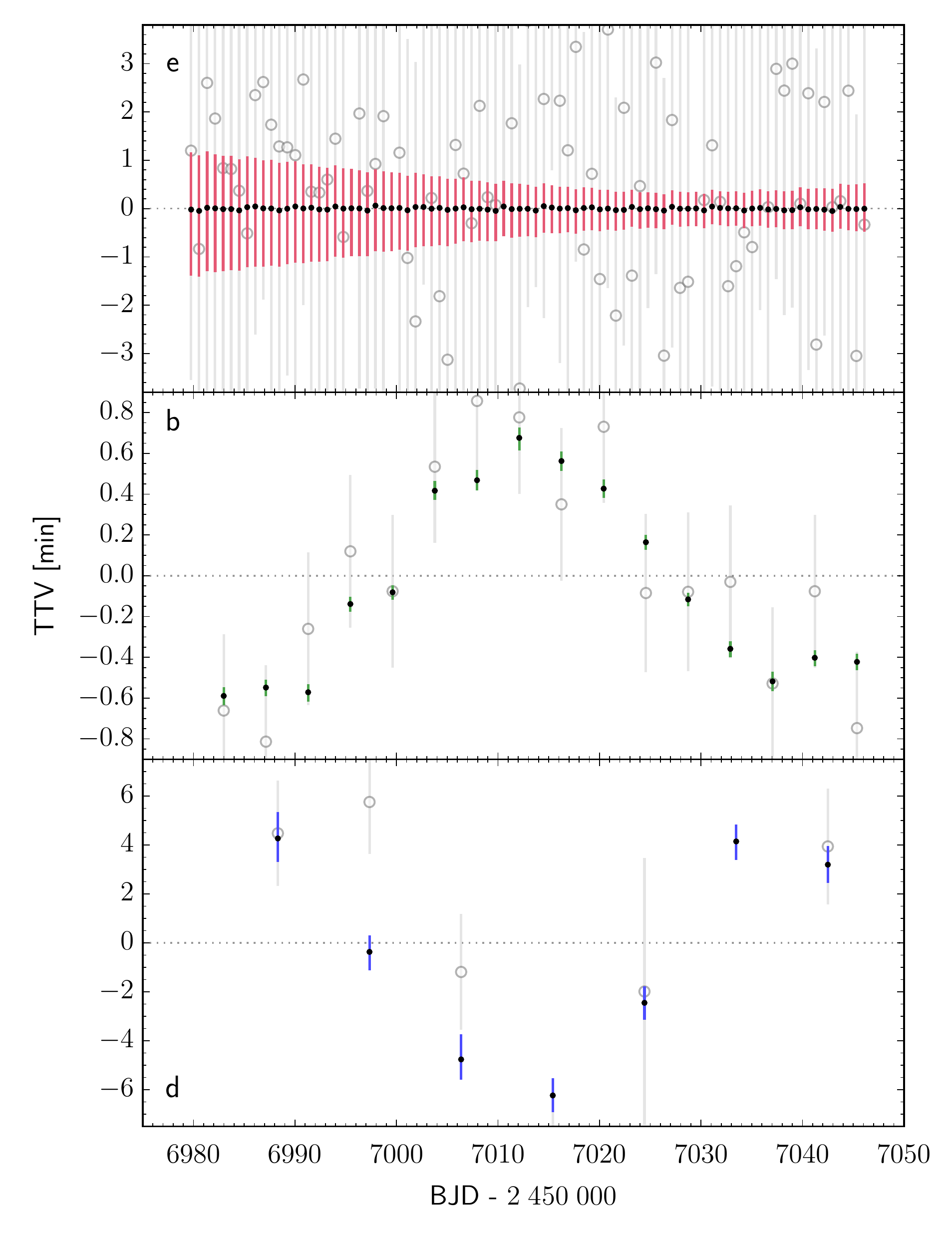}
  \caption{Posterior TTV of planets~e, b, and d (from top to bottom). Black dots with coloured error bars are the posterior TTV from the analysis (median and 68.3\% confidence interval). The grey open circles are the \citet{becker2015} TTV, measured individually from each observed transit. The gain in the mean transit times precision is a factor 9 for planets~b and e, and 3 for planet~d.}
  \label{fig.TTVs}
\end{figure*}

\begin{figure*}
  \hspace{-2.5cm}\includegraphics[height=23cm]{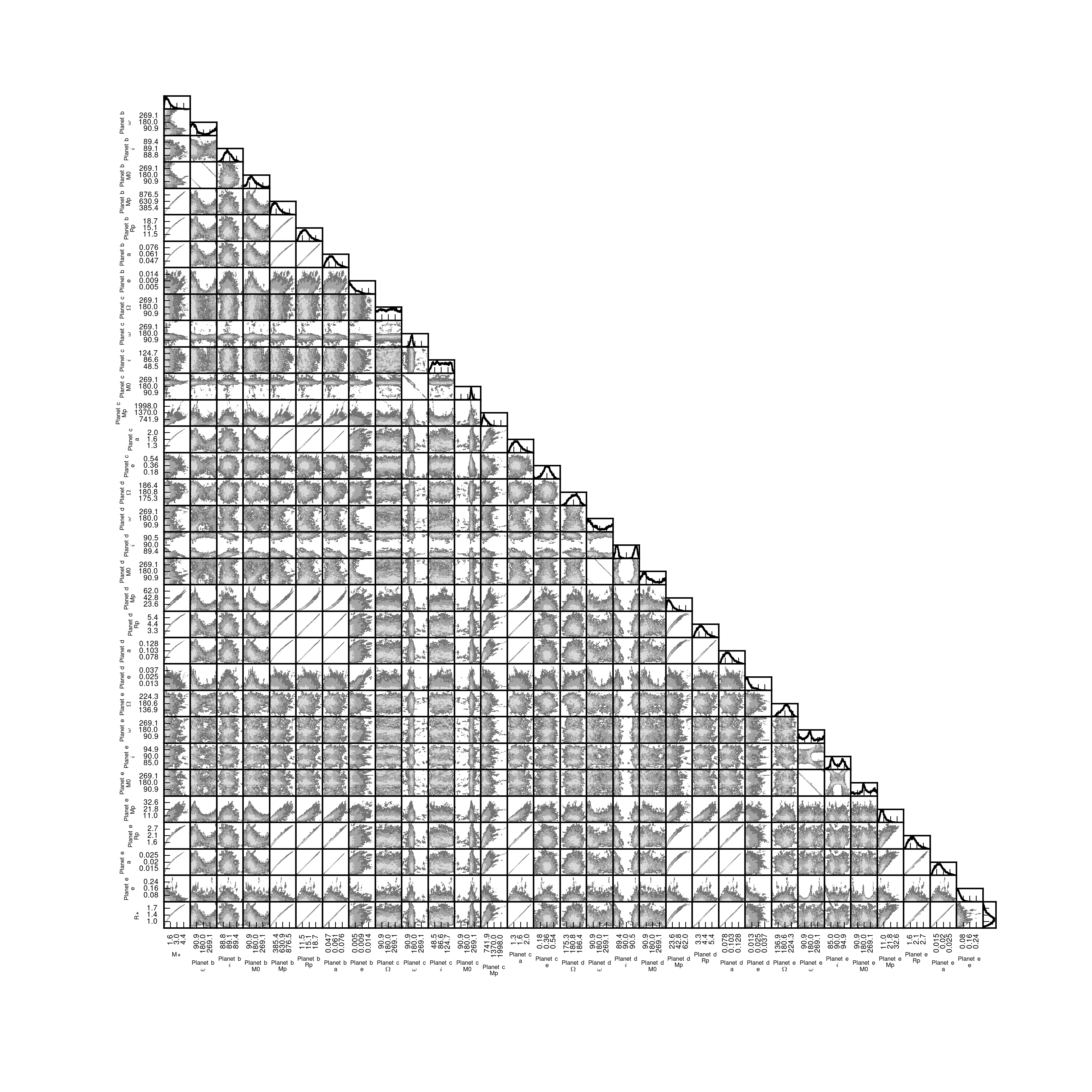}
    \vspace{-2cm}
  \caption{Two-parameter joint posterior distributions for the most relevant MCMC model parameters. The 39.3, 86.5, and 98.9\% two-variable joint confidence regions (in the case of a Gaussian posterior, these regions project on to the one-dimensional 1, 2, and 3~$\sigma$ intervals) are denoted by three different grey levels. The histogram of each parameter is shown at the top of each column, except for the parameter on the last line, which is shown at the end of the line. Units are the same as in Table~\ref{table.results}.}
  \label{fig.pyramid}
\end{figure*}

\begin{figure*}
  \centering
  \includegraphics[height=12cm]{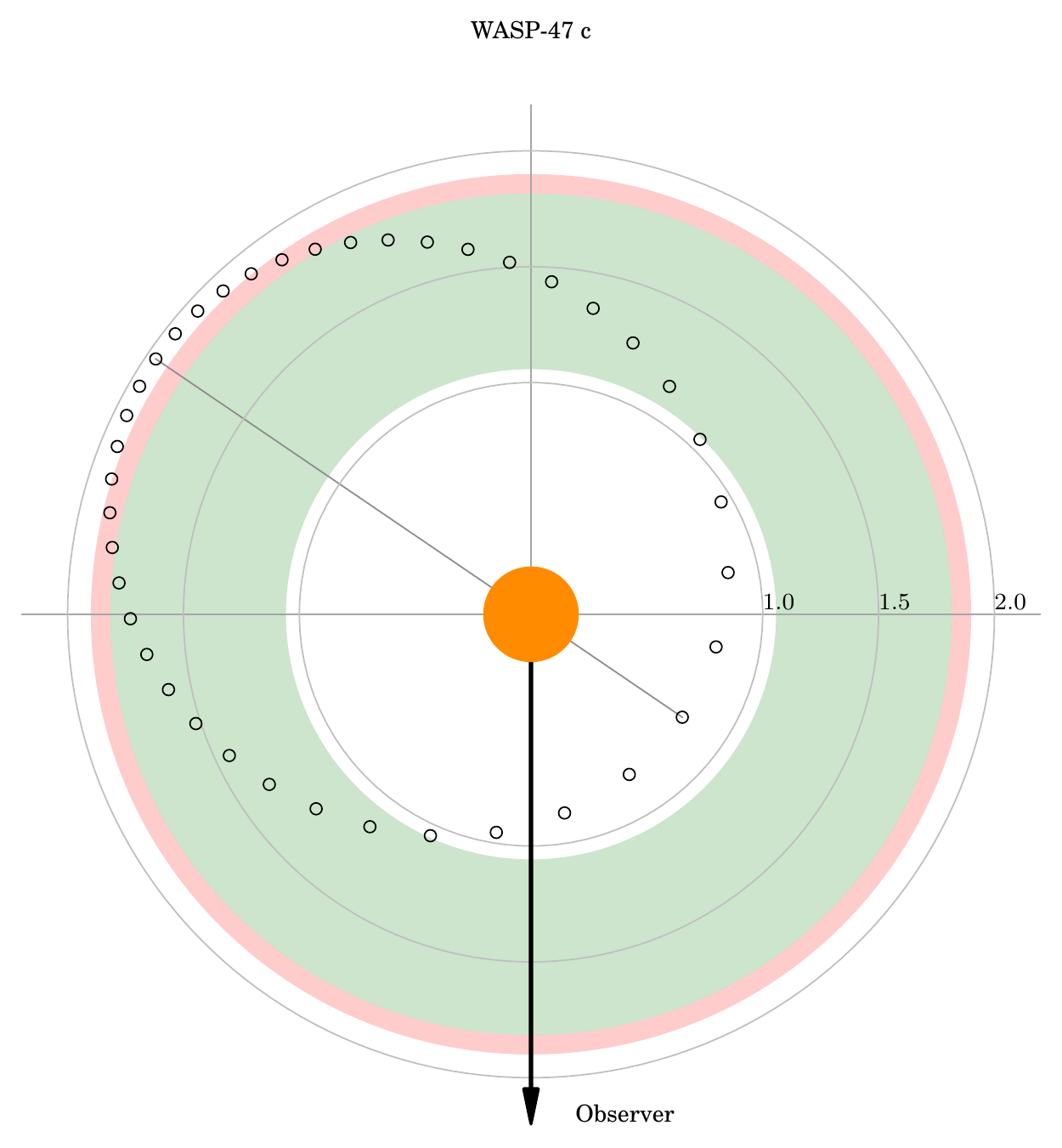}
  \caption{Schematic view of the orbital plane of WASP-47 c. The host star is represented by the orange circle at the centre. The maximum a posteriori orbit is indicated by the empty black points, which are equally spaced in time over the orbit. The orbital movement is counter-clockwise. The angles are measured counter-clockwise from the negative x-axis. The semi-major axis of the orbit is shown as a thin grey line, and the concentric circles are labelled in astronomical units. The black thick arrow points towards the observer. The filled green area is the habitable zone comprised between the runaway greenhouse limit and the maximum greenhouse limit, according to the model of \citet{kopparapu2013}. The red area corresponds to the increased habitable zone if the outer edge is defined by the early-Mars limit.} \label{fig.HZ}
\end{figure*}

\end{appendix}

\end{document}